\documentclass[onecolumn,pre,amsfont,amssymb,amsmath,superscriptadress, nofootinbib]{revtex4-2}
\usepackage{graphicx}
\usepackage[normalem]{ulem}
\usepackage{mathtools}
\usepackage{amsmath}
\usepackage{bm}
\usepackage{gensymb}
\usepackage{hhline}
\usepackage{natbib}
\usepackage[dvipsnames]{xcolor}
\usepackage{hyperref}
\usepackage{duckuments}

\hypersetup{
    colorlinks,
    linkcolor={red!70!black},
    citecolor={blue!70!black},
    urlcolor={blue!70!black}
}
\newcommand{\hbf}[1]{\hat{\mathbf{#1}}}

\begin{document}

\title{Don't look back: Ordering and defect cloaking in non-reciprocal lattice {XY} models}

\author{Pankaj Popli}
\affiliation{Centre for Condensed Matter Theory, Department of Physics, Indian Institute of Science, Bengaluru 560 012,  India
}
\email{pnkjpopli@gmail.com }
\author{Ananyo Maitra}
\affiliation{{Laboratoire de Physique Th\'eorique et Mod\'elisation, CNRS UMR 8089,
		CY Cergy Paris Universit\'e, F-95032 Cergy-Pontoise Cedex, France}}
\affiliation{Laboratoire Jean Perrin, Sorbonne Universit\'{e} and CNRS, F-75005, Paris, France}
\email{nyomaitra07@gmail.com}
\author{Sriram Ramaswamy}
\affiliation{
Centre for Condensed Matter Theory, Department of Physics, Indian Institute of Science, Bengaluru 560 012,  India} 
\affiliation{International Centre for Theoretical Sciences, Tata Institute of Fundamental Research, Bengaluru 560 089, India}
\email{sriram@iisc.ac.in}

\date{\today}
\begin{abstract}
We present a detailed analytical and numerical examination{, on square and triangular lattices, of the} non-reciprocal {planar} {spin model 
introduced} in Dadhichi et al., \href{https://doi.org/10.1103/PhysRevE.101.052601}{Phys. Rev. E {\bf 101}, 052601 (2020)}. We show that the effect of lattice anisotropy should persist at large scales, leading to a ``mass'' for the angle field of the spins, and behaviour not in the ``Malthusian Toner-Tu'' universality class. Numerically, we find evidence of this mass at large values of our non-reciprocity parameter; for smaller values, we find power-law scaling of long-wavelength equal-time correlators in the polar-ordered phase of our lattice model {over the system sizes and wavenumber range explored.}
Focussing on topological defects, we {show numerically} that defect interactions are highly anisotropic with respect to the mean ordering direction. In particular, the constituents of a $\pm 1$ pair are shielded from each other in a class of configurations, deferring their annihilation and allowing time for the nucleation of further defects. The result, we show numerically, is the destruction of the polarised phase via an aster apocalypse reminiscent of that found by Besse et al.,  \href{https://doi.org/10.1103/PhysRevLett.129.268003}{Phys. Rev. Lett. {\bf 129}, 268003 (2022)}, for the Malthusian Toner-Tu equation.
\end{abstract}

\maketitle
\section{Introduction}
The active nature of the Vicsek model \cite{Vicsek}, in which the \textit{aligning} interaction of spins is as in the standard equilibrium XY model, lies in the \textit{movement} of the spins in the direction in which they point \cite{Dadam_rev, Ramaswamy_annu_rev,SR_Aditi_RMP, Toner_Tu_Ramaswamy, Toner_book}. Indeed, motile active units have become a mainstay of research in nonequilibrium statistical mechanics. Motility has been shown to induce phase separation in models without attractive positional interaction \cite{MIPS1, MIPS2} {and to lead \cite{Vicsek,Toner1998, Toner2012, Toner1995, Toner_Tu_Ramaswamy}} to long-range orientationally ordered states even in two dimensions. However, it remains important to disentangle which nonequilibrium phenomena require motility qua motility and which {don't}. For instance, while motion is certainly necessary for motility-induced phase separation, it was {argued in \cite{CLT1, CLT2} and explicitly} shown in \cite{dadhichi}  that it is inessential for advective transport of fluctuations. 
A model of {non-motile} spins with aligning torque on a given spin due to a neighbour depending on the angle between the focal spin and the vector joining it to its neighbour, 
belongs to the Malthusian Toner-Tu universality {class 
and}, as such, displays long-range orientational order. Since then, a number of articles \cite{Loos2022a, Loos2024, Rouzaire2024} have explored related nonreciprocal lattice models numerically and reported a variety of somewhat puzzling results regarding ordering in these systems.

In this---hopefully clarificatory---article, we perform numerical simulations of the inertialess version of the spin model constructed in \cite{dadhichi} and {examine critically} the properties of both our model and those discussed in \cite{Loos2022a, Rouzaire2024}. We show analytically that the spin-wave theory for these lattice models \emph{cannot} belong to the Malthusian Toner-Tu universality class because of residual relevant effects of lattice anisotropy. While this effect can be eliminated---for instance, by using the strategy proposed in \cite{CLT1, CLT2}---we do not do so here. However, the coupling to the lattice anisotropy \emph{cannot} take the model to the \textit{equilibrium} clock model universality class; instead it should lie in a universality class corresponding to the ``Malthusian'' \cite{Toner2012} version of the active clock model \cite{Toner_clock}. As argued in \cite{Toner_clock}, this should imply that a \textit{quasi}-long-range ordered phase \emph{does not} exist for any lattice coordination number, in contrast to passive systems \cite{JKKN}. We then focus on the dynamics of topological defects in this lattice model, on which the effects of lattice anisotropy should be even greater than on spin waves. 

First, even in a hypothetical isotropic system (i.e., one without the lattice anisotropy), while the large-scale structure of the orientation field is universal and can be correctly described within a hydrodynamic theory, the mobility of a defect {cannot. 
Mobility is intrinsically model-dependent as it involves the \emph{core} of the defect, which harbours strong gradients of the order parameter \emph{amplitude}. The value of the mobility 
depends on terms with arbitrary number of gradients of fields, and cannot be obtained from a strictly hydrodynamic theory
\footnote{This issue poses no complications for the thermal equilibrium phase diagram, as the static statistical properties of a state with defects are independent of the mobilities.}.} 

This implies that defect dynamics in active systems is intrinsically more model-dependent than in passive systems. Further, since $-1$ defects in polar active systems stretch out into {walls} \cite{Besse2022} (also, see below), and since these localised structures support large gradients of the angle field, they should be significantly affected by the lattice anisotropy. Nevertheless, in common with \cite{Rouzaire2024}, we find that defect annihilation is highly anisotropic: the time for merging of two defects depends dramatically on their relative
orientation. Because of this slow merging, and the consequent long-lived regions with large gradients, the regions with quasi-static defect pairs serve as nucleation sites for new defects. We believe that this forms the initial stage of the ``aster explosion'' described in \cite{Besse2022} {and \cite{Solon_comm} (which simulated the model discussed here and in \cite{dadhichi})}.

\section{The model}
\begin{figure}[h]
\centering
\includegraphics[scale=0.25]{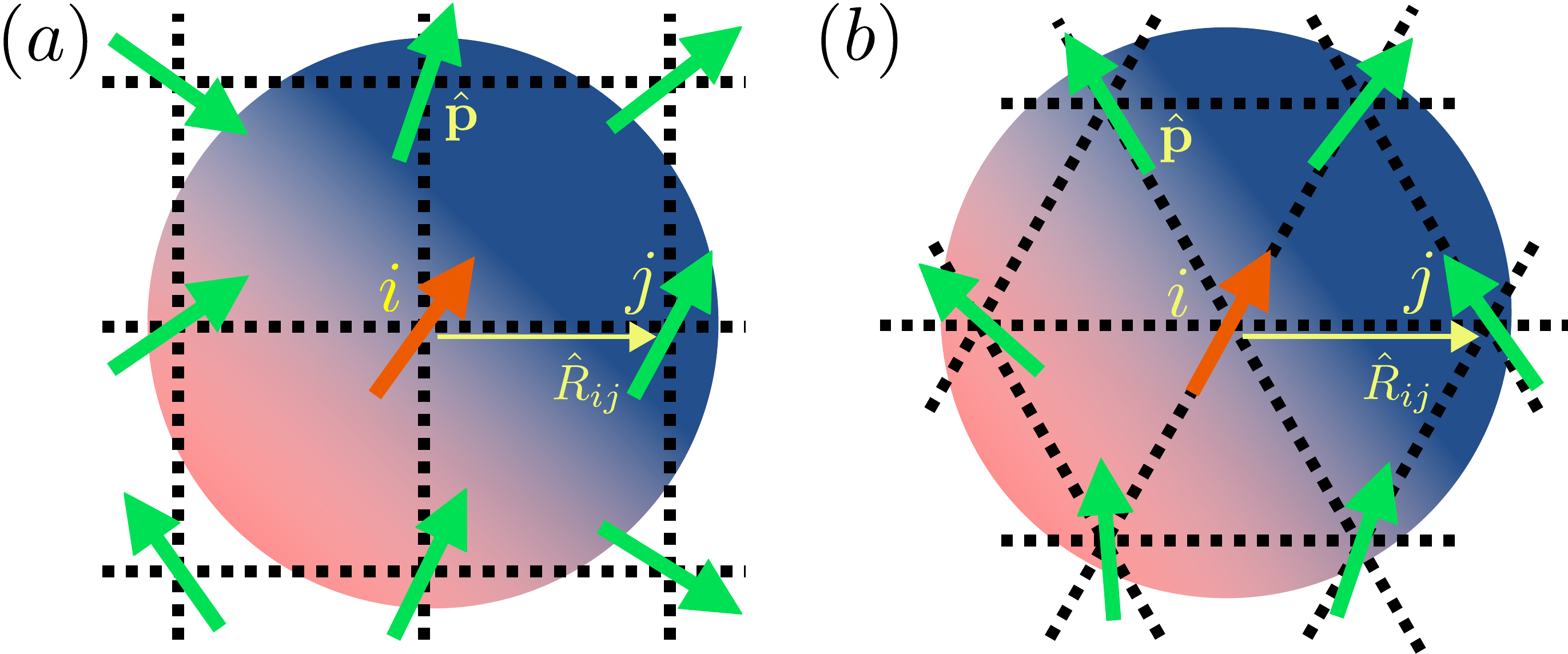}
 \caption{Schematic of our lattice model. The arrows show spins with orientation $\hat{{\bf p}}^i$ at site $i$ interacting with their nearest neighbours on a square {(a)} or a triangular {(b)} lattice. Unlike in equilibrium XY models, the coupling $J^{ij}$ governing the effect of a spin $j$ on the focal spin $i$ (shown in orange) depends on $\hat{{\bf p}}^i\cdot\hbf{R}_{ij}$, where $\hbf{R}_{ij}$ is the lattice vector from spin $i$ to spin $j$; this is schematically represented by the coloured disc around the central spin. This interaction is a continuous variant of a vision cone \cite{Loos2022a, Loos2024, Fernando_vis1, Fernando_vis2}. In common with \cite{Loos2022a, Loos2024, Rouzaire2024} the interaction in our model, first introduced in \cite{dadhichi, dadhichi_arxiv}, is purely orientational, with immotile  spins permanently assigned to lattice sites. This is distinct from \cite{Fernando_vis1, Fernando_vis2} where motile units move towards the neighbours in front of them and have no microscopic aligning interaction.}
\label{fig:schematic}
\end{figure}
We start by describing our lattice model which was first discussed\footnote{An inertial version of the model was constructed in \cite{dadhichi}, and the inertialess version was discussed as a special case. Here, we start with the inertialess version of the model.} in \cite{dadhichi}. An active unit whose orientation or ``spin'' is denoted by a vector $\hbf{p}^i = (\cos \theta^i,\sin \theta^i)$ in real space is permanently associated with the lattice site $i$ which has a position vector $\mathbf{R}^i$.
Each spin $\theta^i$ evolves in time as
\begin{equation}
\dot\theta^i=\sum _{j\in~\mathcal{N}(i)} J^{ij} \sin\left(\theta^j-\theta^i\right)\\+\sqrt{2 T}\mathbf{\zeta}^i\,,
\label{eq:lattice-model}
\end{equation}
where $\mathcal{N}(i)$ denotes the neighbourhood of the lattice site $i$ and $\zeta_i$ is a Gaussian white noise with zero mean and unit strength. The interaction strength $J^{ij}$, in general, depends on $\hbf{p}^i\cdot\hbf{R}^{ij}$. We focus on the specific simple choice $J^{ij}=J+\mathcal{A}~\hbf{p}^i\cdot\hbf{R}^{ij}$ that has two pieces: an XY model-like aligning piece $J$ and a \emph{non-reciprocal} piece that explicitly depends on $\hbf{p}^i$ and the 
{vector connecting the $i$-th and 
$j$-th lattice sites.} 
The form of the latter is such that it cannot arise in a system whose dynamics is governed by an energy function. In the passive limit, $\mathcal{A}=0$, \eqref{eq:lattice-model} reduces to the standard  XY model, with the dynamics arising from an effective Hamiltonian, resulting in perfectly reciprocal interactions. Besides breaking time-reversal symmetry, $\mathcal{A}$ breaks a spatial symmetry as well: in the absence of $\mathcal{A}$, \eqref{eq:lattice-model} is invariant under \emph{independent} rotations of the lattice and the spins. Therefore, the fact that the \emph{lattice} is only invariant under discrete rotations doesn't affect simulations of XY models or our model in the $\mathcal{A}=0$ limit. However, due to $\mathcal{A}$, \eqref{eq:lattice-model} is only invariant under the \emph{joint} rotation of the lattice and the spins. We will therefore see that lattice anisotropy affects the large-scale, long-time behaviour of our model\footnote{{Similarly, lattice anisotropy would affect the hydrodynamic behaviour of potential lattice models of \textit{equilibrium} nematics that can have more than one Frank constant---i.e., are only invariant under the joint rotation of the lattice and the spins---in \emph{all} dimensions. As a result of this, we expect a numerical exploration of such lattice models to reveal a mass for the director field in these cases, at low enough noise strengths. At intermediate noise strengths in two dimensions, we however expect a {critical phase with quasi-long-range order} in which the lattice anisotropy is irrelevant, but the topological defects remain bound \cite{JKKN}. No such regime exists for $d>2$ \cite{Toner_clock}.}}. 

\section{The ordered state: Numerics}
\begin{figure}[h]
\centering
\hspace{-0.5cm}
 \includegraphics[scale=0.31]{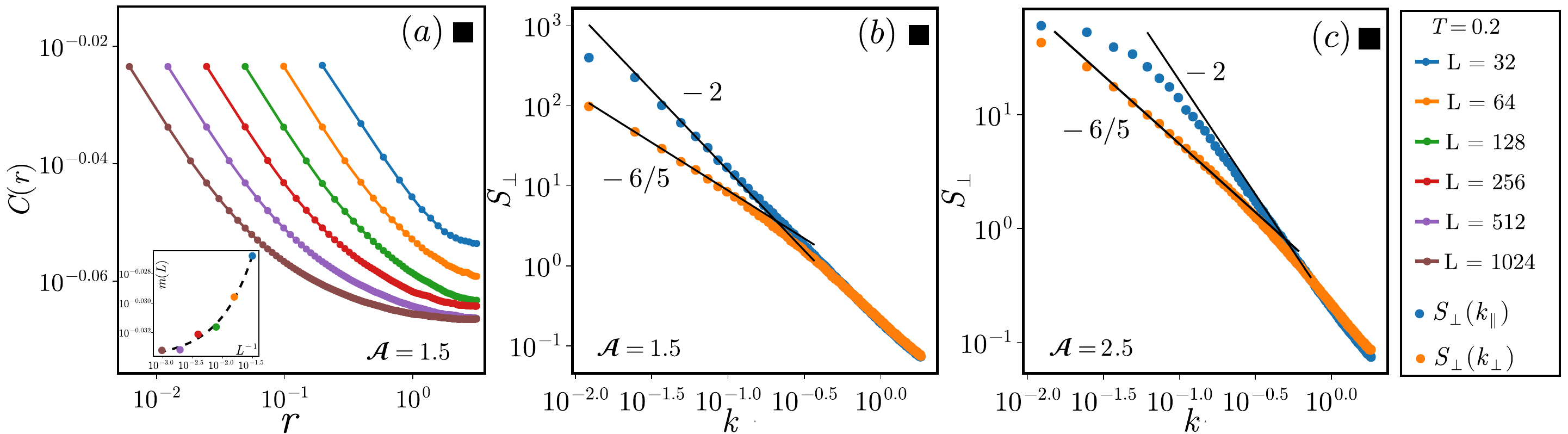}
\caption{
{Statistical properties of the ordered phase on a square lattice. (a) Two-point correlation function $C(r)$ for various system sizes $L$, and $\mathcal{A}=1.5$ with distance $r$ in units of $L/2\pi$, approaches a constant nonzero value for large $r$, indicating long-range order. Inset shows magnetization $m(L)\equiv \sqrt{ C(r=L)}$ as a function of $L^{-1}$. (b) Measured structure factors $S_{\perp}$ as functions of $k_\parallel$ and $k_\perp$ for $L=512$ and $\mathcal{A}=1.5$ show {small-$k$ dependences consistent with 
$k_{\perp}^{-6/5}$ (and $\sim k_{\parallel}^{-2}$ for all but the smallest $k_{\parallel}$)}. 
}
{(c)} Measured structure factors $S_{\perp}$ as functions of $k_\parallel$ and $k_\perp$ for $L=512$ and nonreciprocity strength $\mathcal{A}=2.5$
show {
$k$ dependences consistent with 
$k_{\perp}^{-6/5}$ {except at the smallest $k_\perp$} (and $\sim k_{\parallel}^{-2}$ 
flattening out at the smallest $k_{\parallel}$)}. 
{The flattening of the structure factor at small $k$ is a signature of the generation of a mass in our model. Since it has to vanish as $\mathcal{A}\to 0$, we expect the mass to be larger for larger $\mathcal{A}$. This expectation is borne out by our observation that the structure factor flattens out at a larger wavenumber as $\mathcal{A}$ is increased. All plots were obtained by averaging over five independent noise realisations and the reduced temperature is $T=0.2$ for each of them.
An analogous figure for numerics on a triangular lattice is displayed in Appendix \ref{app:struct}.}
}
\label{fig:correlations}
\end{figure}

We first examine the orientationally ordered state of our model. 
 Initialising our system in an ordered state, we measure 
 isotropically averaged spatial spin correlations $C(r)=\langle{\bf p}(r)\cdot{\bf p}(0)\rangle_{t}$ 
 the angular brackets with a subscript denotes an average over that variable,  for different system sizes $L$, nonreciprocity $\mathcal{A}$,  and noise strength $T$.  
 We benchmark our code by measuring these quantities for $\mathcal{A}=0$ which, as expected, displays quasi-long-range order {below a defect-unbinding transition at temperature $T_{KT} \simeq 0.8$ for a $128 \times 128$ square lattice.}. {For $\mathcal{A}\neq0$, we find that $C(r)$ has a non-zero value at large $r$ signalling the existence of long-range order. That, in itself, is not too surprising: as we will discuss in the next section, when the order parameter space and the real space coincide, as it does in our model for $\mathcal{A}\neq 0$, lattice anisotropy explicitly breaks the continuous rotation symmetry of the model to a discrete one, even in equilibrium. Therefore, at low temperatures, we would expect long-range order in any such model. However, {the reduction to a discrete symmetry would imply 
 that the static structure factor of angular fluctuations 
 $S_\perp({\bf k})\equiv \langle|\hbf{p}_{\perp}({\bf k},t)|^2\rangle$,
 where $\hbf{p}_{\perp}$ is the component of $\hbf{p}$ perpendicular to the instantaneous macroscopic ordering direction and ${\bf k}$ is the wavevector, should go} to a $|{\bf k}|-$independent constant at small $|{\bf k}|$. We plot $S_\perp(k_\parallel)$ and $S_\perp(k_\perp)$, where $k_\parallel$ and $k_\perp$ are the components of ${\bf k}$ along and orthogonal to the ordering direction respectively, {in Fig.~\ref{fig:correlations}b and c, for two values of $\mathcal{A}$}{. While {for $\mathcal{A}=1.5$,} we find a clear departure
 from the $1/k_\perp^2$} 
 expected when $\mathcal{A}=0$, they {show a dependence on wavenumber consistent with a power law} at least in the regime we have explored. The ``mass'' (see the next section) conferred on the transverse fluctuations by the identification of the order parameter and lattice directions is not detected at these scales {at this value of $\mathcal{A}$}. In fact, Fig.~\ref{fig:correlations}b is consistent with the findings of \cite{Besse2022}.  
 }
{In contrast, a distinct flattening of the structure factor at small $k_\parallel$ is observed in systems with $\mathcal{A}=2.5$ (to measure this structure factor, we only averaged over realisations in which the ordered phase was not destroyed via an aster explosion---to be discussed below---within the run-time of our simulation). While the flattening is less obvious at small $k_\perp$, this is a clear signature of the generation of mass. We discuss the mechanism for this in the next section.}

\section{From lattice to continuum: the lingering effects of lattice anisotropy}
In this section, we discuss whether the ordered state of our model---and related models from other groups \cite{Loos2022a, Loos2024,Rouzaire2024}---belong to the Malthusian Toner-Tu universality class\footnote{whose exponents are unknown; see \cite{AM_malthus}}. In particular, we discuss the effects of the lattice anisotropy on the long-time, large-scale physics of a state ordered along $\hbf{x}$. We begin by rewriting \eqref{eq:lattice-model} as a dynamics of $\hbf{p}_i$:
\begin{equation}
d\hbf{p}^i/dt= \sum _{j\in~\mathcal{N}(i)} \left( J+\mathcal{A}~\hbf{p}^i\cdot\hbf{R}^{ij}\right) \hbf{p}^i \times \hbf{p}^j\times\hbf{p}^i+\sqrt{2 T} \hbf{p}^i \times \boldsymbol{\zeta}^i\times\hbf{p}^i\,,
\label{eq:interaction-terms}
\end{equation}
where $\boldsymbol{\zeta}^i$ is a two-component vector of zero-mean, unit variance Gaussian white noises whose components are uncorrelated.
For defect-free regions\footnote{In general, we must implement a coarse-graining procedure which results in a smooth \textit{non-unit} vector field $\mathbf{p}$, from which we can construct an equation for the director $\hbf{p}$.}, we can define a smooth unit vector \emph{field} $\hbf{p}$ such that $\hbf{p}({\bf x}_i,t)\equiv\hbf{p}^i$, where ${\bf x}_i$ is the spatial position of lattice site $i$, {by assuming $\hbf{p}_j$ does not vary much from site to site and
expanding} 
\begin{equation}
\hbf{p}_\alpha({\bf x}_j,t)
\approx \hbf{p}_\alpha({\bf x}_i,t) + \hbf{{R}}^{ij}_{\beta}\partial_{\beta}\hbf{p}_\alpha({\bf x}_i,t)
+\frac{1}{2}\hbf{{R}}^{ij}_{\beta}\hbf{{R}}^{ij}_{\gamma}\partial_\beta\partial_{\gamma}\hbf{p}_\alpha({\bf x}_i,t)+\frac{1}{6}\hbf{{R}}^{ij}_{\beta}\hbf{{R}}^{ij}_{\gamma}\hbf{{R}}^{ij}_{\mu}\partial_\gamma\partial_\beta\partial_{\mu}\hbf{p}_\alpha({\bf x}_i,t)
+\dots
\label{expan}
\end{equation}
Here, Latin letters label lattice sites and Greek letters represent spatial components.
Because our system is on a lattice, $\hbf{R}_{ij}$ cannot assume arbitrary values but must correspond to the lattice vectors. For a square lattice, these are ${\bf e}^a\equiv\{\pm(1,0), \pm(0,1)\}$. For a triangular lattice, ${\mathbf{e}}^a\equiv\{\pm(1,0), \pm(\pm 1/2,\sqrt{3}/2)\}$.
Using \eqref{expan} and retaining terms only up to first order in gradients, we calculate the effect of $\mathcal{A}$ in the ordered state:
\begin{equation}
\left(\sum _{j\in~\mathcal{N}(i)} \mathcal{A}~\hbf{p}^i\cdot\hbf{R}^{ij}\hbf{p}^i \times \hbf{p}^j\times\hbf{p}^i\right)_\alpha\approx\mathcal{A}\delta^T_{\alpha\beta}\sum_{a}\hbf{p}_\mu({\bf x}_i,t){\bf e}_\mu^a {\bf e}_\gamma^a\partial_{\gamma}\hbf{p}_\beta({\bf x}_i,t)\,,
\end{equation}
where $\delta^T_{\alpha\beta}=\delta_{\alpha\beta}-\hbf{p}_\alpha\hbf{p}_\beta$.
For a square lattice, $\sum_{a}{e}_\beta^a {e}_\alpha^a=2\delta_{\alpha\beta}$ and for a triangular lattice, $\sum_{a}{e}_\beta^a {e}_\alpha^a=3\delta_{\alpha\beta}$. Thus, at this order in gradients,
\begin{equation}
\label{eq:fixed-length_advection}
\left(\sum _{j\in~\mathcal{N}(i)} \mathcal{A}~\hbf{p}^i\cdot\hbf{R}^{ij}\hbf{p}^i \times \hbf{p}^j\times\hbf{p}^i\right)_\alpha\approx \mathcal{A}c\delta^T_{\alpha\beta}\hbf{p}_\gamma({\bf x}_i,t)\partial_{\gamma}\hbf{p}_\beta({\bf x}_i,t)i=\mathcal{A}c\hbf{p}_\gamma({\bf x}_i,t)\partial_{\gamma}\hbf{p}_\beta({\bf x}_i,t)\,,
\end{equation}
where $c$ is a numerical constant that depends on the coordination number and the second equality arises because $\hbf{p}({\bf x},t)$ is a unit vector and $\hbf{p}_\beta({\bf x},t)\partial_\gamma\hbf{p}_\beta({\bf x},t)=0$. Using the same argument and noting that $\sum_a{\bf e}^a=0$, we get
\begin{equation}
\label{eq:fixed-length_diffusion}
\left(\sum _{j\in~\mathcal{N}(i)} J\hbf{p}^i \times \hbf{p}^j\times\hbf{p}^i\right)_\alpha\approx J\frac{c}{2}\delta^T_{\alpha\beta}\nabla^2\hat{p}_\beta({\bf x}_i,t)\,.
\end{equation}
Therefore, to $\mathcal{O}(\nabla^2)$, a continuum version of \eqref{eq:interaction-terms} reads
\begin{equation}
    \partial_t\hat{p}_\alpha=c\left(\frac{J}{2}\delta^T_{\alpha\beta}\nabla^2\hat{p}_\beta+\mathcal{A}\hat{p}_\beta\partial_\beta\hat{p}_\alpha\right)+\sqrt{2T}\delta^T_{\alpha\beta}{\zeta}_\beta\,
    \label{cont_sq}
\end{equation}
{ignoring \textit{reciprocal} consequences of the $\mathcal{A}$ term  that arise in \cite{dadhichi} when \eqref{eq:fixed-length_advection} is developed to order $\nabla^2$ and lead to anisotropic diffusion}. Eq. \eqref{cont_sq} is the fixed-length version of the density-free Toner-Tu model 
\cite{Toner2012,AM_malthus} 
with the non-reciprocity endowing the \emph{polarisation} equation with an effective self-advective nonlinearity. At this stage, our discussion seems to suggest that {our} lattice model belongs to the 
universality class of \cite{Toner2012}, at least as far as the spin-wave theory is concerned. We will explore this suggestion further below. However, before that, some comments about certain analogous lattice models published following \cite{dadhichi} are in order.

Loos et al. \cite{Loos2022a,Loos2024} considered a non-reciprocal lattice model with a strict vision cone: instead of the $\mathcal{A}$ interaction we use, they assumed an interaction that is $0$ for $\hbf{p}^i\cdot\hbf{R}^{ij}>\phi_c$, where $\phi_c$ is a critical angle whose value is an additional model parameter and non-zero otherwise. It is difficult to obtain the continuum equations that the ordered phase of a model with such a hard cut-off would obey. Recently, Rouzaire et al. \cite{Rouzaire2024} constructed a similar model with a smooth cut-off. They consider an interaction kernel which yields $\partial_t\hbf{p}_i\propto \sum _{j\in~\mathcal{N}(i)}Je^{\sigma\hbf{p}^i\cdot\hbf{R}^{ij}}\hbf{p}^i \times \hbf{p}^j\times\hbf{p}^i$. For small $\sigma$, we can expand
\begin{equation}
e^{\sigma\hbf{p}^i\cdot\hbf{R}^{ij}}=1+\sigma\hbf{p}^i\cdot\hbf{R}^{ij}+\frac{\sigma^2}{2}(\hbf{p}^i\cdot\hbf{R}^{ij})^2+\frac{\sigma^3}{6}(\hbf{p}^i\cdot\hbf{R}^{ij})^3+\frac{\sigma^4}{24}(\hbf{p}^i\cdot\hbf{R}^{ij})^4+\frac{\sigma^5}{120}(\hbf{p}^i\cdot\hbf{R}^{ij})^5+...
\end{equation}
When only terms up to $\mathcal{O}(\sigma)$ are retained, this interaction therefore corresponds to the one in \eqref{eq:interaction-terms} with $\sigma=\mathcal{A}/J$. We now discuss the effect of the terms at higher order in $\sigma$ on the physics of the ordered states. We will discuss the results for a square lattice in the main text and present the ones for a triangular lattice in the appendix~\ref{A:angle_field_triangular}. Using \eqref{expan}, and expanding to second order in gradients and fifth order in $\sigma$, we get 
\begin{multline}
   \left(\sum _{j\in~\mathcal{N}(i)} Je^{\sigma\hbf{p}^i\cdot\hbf{R}^{ij}}\hbf{p}^i \times \hbf{p}^j\times\hbf{p}^i\right)_\alpha\approx J\delta_{\alpha\beta}^T\sum_a\Bigg[\frac{1}{2}{\bf e}_\mu^a {\bf e}_\gamma^a\partial_\mu\partial_{\gamma}\hbf{p}_\beta({\bf x}_i,t)+\sigma\hbf{p}_\mu({\bf x}_i,t){\bf e}_\mu^a {\bf e}_\gamma^a\partial_{\gamma}\hbf{p}_\beta({\bf x}_i,t)\\+\frac{\sigma^2}{4}\hbf{p}_\mu({\bf x}_i,t)\hbf{p}_\nu({\bf x}_i,t){\bf e}_\mu^a{\bf e}_\nu^a{\bf e}_\lambda^a {\bf e}_\gamma^a\partial_\lambda\partial_{\gamma}\hbf{p}_\beta({\bf x}_i,t)+\frac{\sigma^3}{6}\hbf{p}_\mu({\bf x}_i,t)\hbf{p}_\nu({\bf x}_i,t)\hbf{p}_\lambda({\bf x}_i,t){\bf e}_\mu^a{\bf e}_\nu^a{\bf e}_\lambda^a {\bf e}_\gamma^a\partial_{\gamma}\hbf{p}_\beta({\bf x}_i,t)
   \Bigg]\,,
\end{multline}
where we have used the fact that the rank $n$ tensor $\sum_a[{\bf e}^a]^{\otimes n}$ vanishes for odd $n$. Writing $\hbf{p}({\bf x},t)\equiv\{\cos\theta({\bf x},t),\sin\theta({\bf x},t)\}$, where we have defined an angle \emph{field} $\theta({\bf x},t)$ whose values at lattice sites $i$ correspond to $\theta^i$, {the equation of motion of \cite{Rouzaire2024} on a square lattice to order $\sigma^3$ reduces to}
\begin{equation}
\partial_t\theta=2J\Bigg[\frac{1}{2}\nabla^2\theta+\sigma(\cos\theta\partial_x\theta+\sin\theta\partial_y\theta)+\frac{\sigma^2}{4}(\cos^2\theta\partial_x^2\theta+\sin^2\theta\partial_y^2\theta)+\frac{\sigma^3}{6}(\cos^3\theta\partial_x\theta+\sin^3\theta\partial_y\theta)
\Bigg]+\sqrt{2T}\xi_\theta\,.
\label{Levis_cont_sq}
\end{equation}
Several comments on \eqref{Levis_cont_sq} are in order. First, in isotropic space, rotation symmetry ensures that the angle field equation is invariant under the transformation
\begin{equation} \label{rotinv} 
\theta\to\theta+\theta_0, \, x\to x\cos\theta_0-y\sin\theta_0, \, y\to y\cos\theta_0+x\sin\theta_0. 
\end{equation} 
Additionally, reflection symmetry ensures that it is invariant
under $y\to -y$, $\theta\to-\theta$ \cite{AM_malthus}.
While \eqref{Levis_cont_sq} is symmetric under these transformations for terms up to $\mathcal{O}(\sigma)$, it is clearly not symmetric under \eqref{rotinv} for terms at higher orders in $\sigma$. For instance, the term $\propto\sigma^3/6$ clearly breaks this symmetry at $\mathcal{O}(\nabla)$ in a way that modifies the scaling behaviour qualitatively (a hyperscaling relation that relates two relevant nonlinearities of the flocking model in isotropic space is no longer valid).
Therefore, the microscopic model proposed in \cite{Rouzaire2024} cannot describe the spin wave dynamics of flocks in a fully rotation-invariant space. Second, one would expect the coupling of the order parameter to lattice directions to lead to an aligning field that favours alignment along any one of those directions, and thus a restoring torque proportional to the deviation $\theta$ of the orientation from that direction \cite{JKKN}. Such a term is apparently absent in the dynamics as derived above, 
but fluctuations may generate it {though the interaction in \cite{Rouzaire2024} doesn't have a special aligning direction, unlike traditional models with crystal fields \cite{JKKN, Toner_clock}. First notice that even though the model is not rotation-invariant, $\theta$ is still a compact, pseudoscalar variable and its equation of motion therefore has to be symmetric under $\theta\to\theta+2\pi$. This implies that the ``mass'' {term}, if it exists, must be $\propto\sin\theta$. We notice that the $\mathcal{O}(\sigma^2)$ term in \eqref{Levis_cont_sq} can be rewritten as $\partial_x^2\theta+\sin^2\theta(\partial_y^2\theta-\partial_x^2\theta)$. In a system that does not have rotation invariance and has long-range order, {$\sin^2\theta(\partial_y^2\theta-\partial_x^2\theta)$ can be replaced in a ``Hartree'' spirit by $\langle \sin\theta(\partial_y^2\theta-\partial_x^2\theta)\rangle\sin\theta$. No symmetry dictates that the quantity in angle brackets should vanish, so it should in general be non-zero, 
imparting} a mass to the orientational fluctuations\footnote{{Note that superficially similar nonlinearities arise in the dynamics of polar and nematic active systems as well; for instance, at $\mathcal{O}(\nabla^2)$, nonlinearities that remain invariant under the transformation \eqref{rotinv} include $\{\sin2\theta[(\partial_x\theta)^2-(\partial_y\theta)^2]-2\cos2\theta\partial_x\theta\partial_y\theta\}$ and $[\cos2\theta(\partial_x^2\theta-\partial_y^2\theta)+2\sin2\theta\partial_x\partial_y\theta]$ with independent coefficients. Naively, it would seem that a term $\propto\sin2\theta$ may be generated from $\sin2\theta[(\partial_x\theta)^2-(\partial_y\theta)^2]$ in a state with long range order. It would also seem that another such mass term may be generated from $\cos2\theta(\partial_x^2\theta-\partial_y^2\theta)$. Of course, no such term is generated because unlike both of these nonlinearities, a term $\propto\sin2\theta$ or $\sin\theta$ is not invariant under \eqref{rotinv}. In contrast, such terms can be generated in \eqref{Levis_cont_sq} because it is not invariant under \eqref{rotinv}.}}.} 
{{In Appendix \ref{app:struct} we} display the static structure factor $S_\perp$ of angular fluctuations obtained numerically using the model of \cite{Rouzaire2024}, which clearly shows that a mass is generated.}
This places the models considered in \cite{Rouzaire2024, Loos2022a, Loos2024} within the active clock model universality class. It was conclusively shown in \cite{Toner_clock} that, unlike passive clock models which at high enough lattice coordination numbers and at finite noise strengths behave as isotropic spin systems, active $q$-state clock models remain only $q$-fold symmetric, and are thus distinct from polar active systems in isotropic space, for any $q$ no matter how large. 
Since order in the presence of anchoring to lattice directions should be long-ranged, the quasi-long-range ordered state reported in \cite{Loos2022a} 
is likely to be a finite-size effect.

Since the interaction in \cite{Rouzaire2024} corresponds to the one we consider here to $\mathcal{O}(\sigma)$, and since the terms we identified as breaking isotropy all arise at higher order in $\sigma$, one might imagine that the spin wave theory of our model belongs to the Malthusian Toner-Tu universality class. We now argue that this is not the case. To do so, we use \eqref{expan} to {obtain the continuum version of our model} \eqref{cont_sq} to $\mathcal{O}(\nabla^4)$ on square lattices{, yielding the equation of motion} 
\begin{multline}
\partial_t\theta=2\Bigg(\mathcal{A}(\cos\theta\partial_x\theta+\sin\theta\partial_y\theta)+\frac{J}{2}\nabla^2\theta+\frac{\mathcal{A}}{6}\{\cos\theta[\partial_x^3\theta-(\partial_x\theta)^3]+\sin\theta[\partial_y^3\theta-(\partial_y\theta)^3]\}\\+\frac{J}{24}\{\partial_x^4\theta+\partial_y^4\theta-6[(\partial_x\theta)^2\partial_y^2\theta+(\partial_y\theta)^2\partial_x^2\theta]\}\Bigg)+\sqrt{2T}\xi_\theta\,
\label{cont_sq4}
\end{multline}
{for the angle field.} The 
terms $\propto\mathcal{A}/6$ and $J/24$ 
in the large 
parentheses are not symmetric under \eqref{rotinv}, for distinct reasons. First, we discuss the term $\propto J/24$ which \emph{is} symmetric under $\theta\to\theta+\theta_0$. This is because space and spin are decoupled in our model when $\mathcal{A}=0$, and spin rotation symmetry is preserved irrespective of the spatial anisotropy. The spatial gradient structure reflects the four-fold anisotropy of the lattice.
The term $\propto\mathcal{A}/6$ is more interesting: the nonlinearities
 $\cos\theta(\partial_x\theta)^3$ and $\sin\theta(\partial_y\theta)^3$ have structures that can renormalise the coefficients of the terms $\cos\theta(\partial_x\theta)$ and $\sin\theta(\partial_y\theta)$ 
{implying} that the coefficient of the terms $\sin\theta\partial_y\theta$ and $\cos\theta\partial_x\theta$ need not remain equal in a renormalised theory, breaking the symmetry of our model under \eqref{rotinv} even at $\mathcal{O}(\nabla)$. Since both of these terms contain relevant nonlinearities \cite{AM_malthus} (with a symmetry-mandated hyperscaling relation keeping the ratio of their coefficients fixed under renormalisation in flocks in isotropic space), the hydrodynamic behaviour of the ordered phase of our model cannot belong to the Toner-Tu universality class. 

{Whether a term $\propto\sin\theta$ is generated in our model is more complicated to resolve analytically. As the dynamics is not symmetric under arbitrary rotations, no symmetry \textit{forbids} such a term, which leads us to expect that it must arise. In all likelihood, it does so through terms of the form $\sin\theta(\partial_y\theta)^3$ where the average $\langle (\partial_y\theta)^3\rangle$ could in principle be nonzero beyond the Gaussian approximation. The fact that our model does have a mass that increases with $\mathcal{A}$ is however borne out by our numerical simulations\footnote{{After our article was uploaded on arXiv, another preprint examining this class of models and reaching similar conclusions regarding the mass generation and the lingering effects of anisotropy on hydrodynamics, appeared \cite{Solon_aniso}}}.}


Our model on a triangular lattice has similar qualitative properties; to see this, however, we have to expand \eqref{expan} to at least $\mathcal{O}(\nabla^5)$. Below this, the effect of lattice anisotropy is not evident in the continuum theory. When expanded to $\mathcal{O}(\nabla^5)$, the term $\propto\mathcal{A}$ doesn't remain invariant under \eqref{rotinv}. Further, these have terms that can renormalise the coefficients of $\sin\theta\partial_y\theta$ and $\cos\theta\partial_x\theta$ differently, making the ordered phases {and ordering transitions} of the model on the triangular and square lattices belong to the {corresponding active clock model universality classes.}

\begin{figure}[h]
\centering
\includegraphics[scale=0.3]{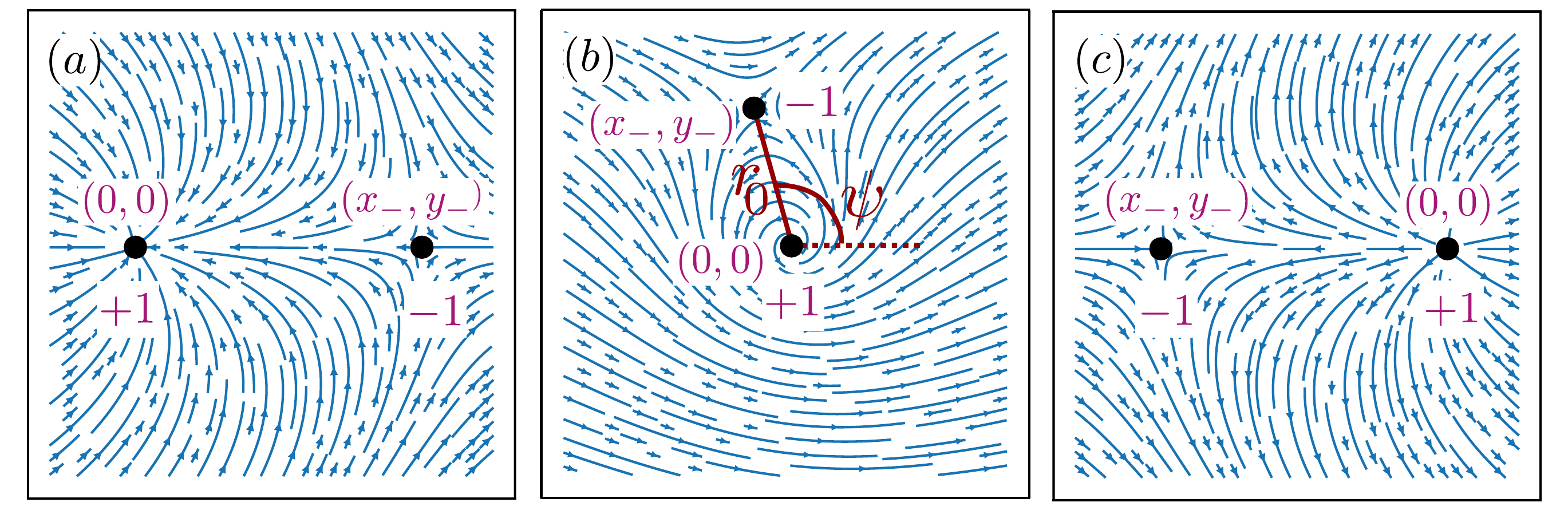}
\caption{{
Typical initial configurations for examining dynamics of $\pm 1$ defect pairs. The vector fields depict $\hbf{p}=(\cos\theta, \sin\theta)$, with $\theta =\tan^{-1}{y}/{x} - \tan^{-1}(y-y_-)/(x-x_-)$ where $(x_-,y_-)$ is the position of the $-1$ defect and the $+1$ defect is placed at the origin. $\psi=0$ in (a), $7\pi/12$ in (b) and $\pi$ in (c).}}
\label{fig:defect-field}
\end{figure}
The central conclusion of this section is that the class of lattice models \cite{Rouzaire2024,Loos2022a,Loos2024} and that defined in this paper, with non-reciprocality as in \eqref{eq:lattice-model}, in which the effect of a spin $\hbf{p}^j$ at site $j$ to a spin $\hbf{p}^i$ at site $i$ depends on the angle between $\hbf{p}^i$ and the lattice direction $\hbf{R}^{ij}$ from $i$ to $j$, remain anisotropic in the continuum limit. The resulting partial differential equations of motion possess only the discrete symmetry of the underlying lattice even at leading order in gradients. For the models of \cite{Rouzaire2024} and \cite{Loos2022a,Loos2024} we show that this implies the presence of a ``mass'' term for the broken-symmetry mode in the ordered phase. We suspect, but have not established, that such a term arises for our model as well. While the effect of lattice anisotropy can be eliminated---for instance, by using the strategy proposed in \cite{CLT1, CLT2, Toner_book}---we do not attempt it here.

\begin{figure}[ht]
\centering
\includegraphics[scale=0.15]{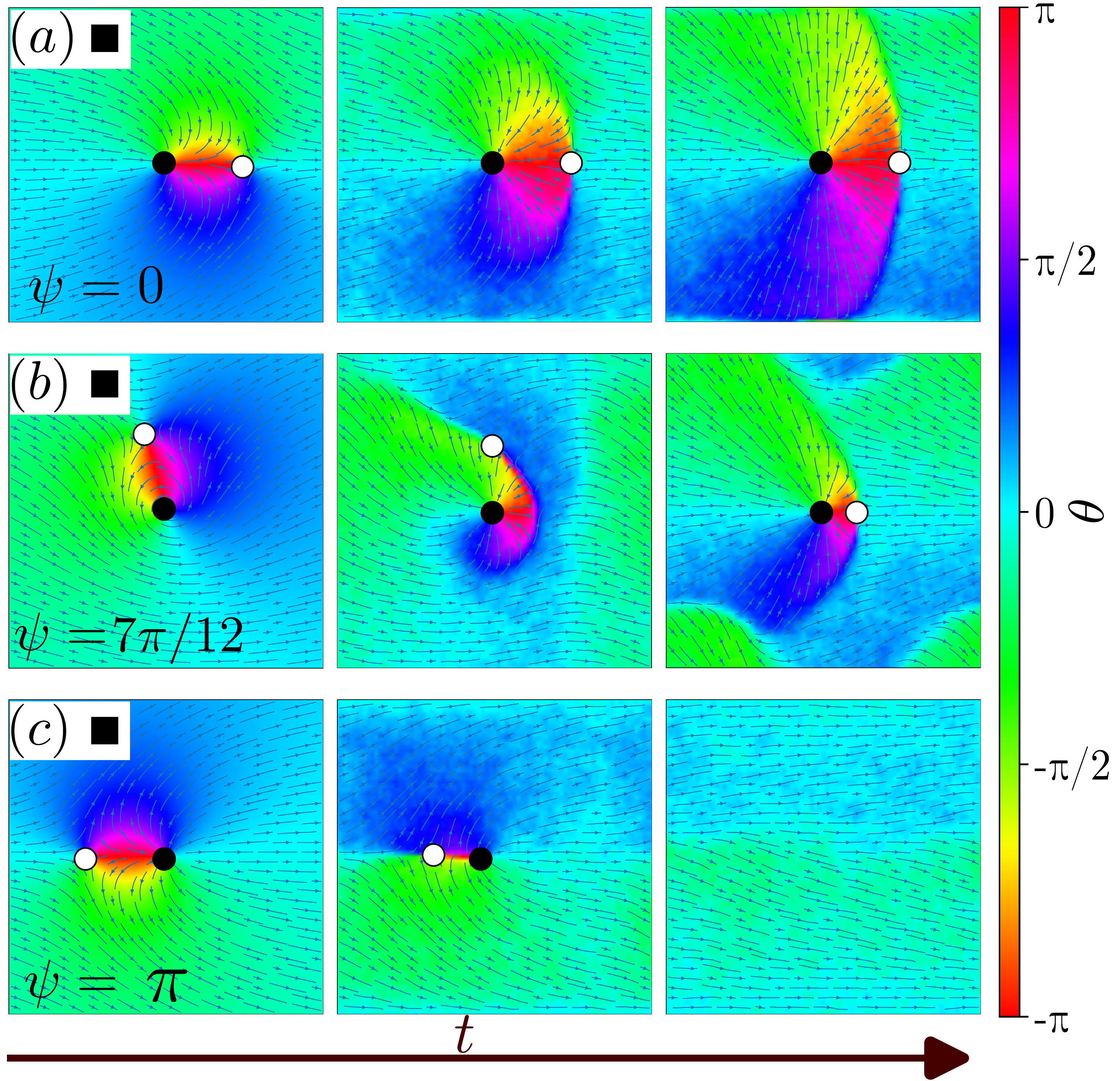}
\caption{{Dynamics of $\pm 1$ defect pairs {(here, and in the rest of the article, the position of a $+1$ defect is marked by a black dot and the position of a $-1$ defect is marked by a white dot)} with initial configurations as shown in} Fig.~\ref{fig:defect-field}. Defect pairs with initial separations of $r_0=10$  were placed in a square lattice of size $L = 64$ with $\mathcal{A}=0.6$ and $T=0.01$. {For {(a)} $\psi=0$, the $\pm 1$ defect pair does not recombine to yield a defect-free state within the timescale of our simulation. Instead, the orientation field around the $-1$ defect has a large-scale wall-like deformation encircling the $+1$ aster; the dynamics in this configuration is so slow that we do not see it evolve within our simulation timescale. {Also see supplementary video SV1.}}
{In {(b)}, initially $\psi=7\pi/12$. In this case, the orientation field again strongly deforms around the $-1$ defect forming a wall-like structure; the defect then moves along this line till $\psi=0$. The final configuration that we observe is analogous to the one in {(a)}. {Also see supplementary video SV2.}} 
{In {(c)}, $\psi=\pi$ and the $\pm 1$ defect pair merges. {Also see supplementary video SV3.}}
}
\label{fig:defect-configs}
\end{figure}
\section{Topological defects: anisotropic cloaking and aster explosion} 
The structure of \eqref{cont_sq} suggests the presence of a length scale $r_c\equiv J/\mathcal{A}$ above which nonreciprocity may be expected to be significant. It is also clear from the form of the advection term that its effect should be \textit{anisotropic}, with a strong asymmetry with respect to the direction of the polarization ${\bf p}$. We now examine how $r_c$ emerges as a key lengthscale in the interactions between topological defects.

In the $\mathcal{A}=0$ limit, Eq. \eqref{eq:lattice-model} corresponds to the standard two-dimensional XY model, for which a topological defect configuration with a charge $n$ placed at the origin is given by $\theta({\bf x}) = n \tan^{-1}({y}/{x}) + \phi$, where $\phi$ is a constant and determines the shape of the charge. For $n= +1$, an anticlockwise or clockwise vortex is produced when $\phi = \pm \pi /2$, while $\phi = 0$ and $\pi$ result respectively in outward- and inward-pointing asters. Any other value of $\phi$ produces a spiral. A $n=-1$ defect has a twofold symmetry and $\phi$ rotates it rigidly. Of course, for the XY model as described, the distinction amongst these three structures is purely notional, as the model is invariant separately under spin and space rotations\footnote{Even in the passive limit, order-parameter and tangent spaces of course coincide for liquid crystals, whether polar \cite{Kung} or apolar \cite{degp}. Such models too are therefore invariant only under joint rotation of the two spaces.}. The difference becomes physical once we introduce nonreciprocity in the form of a dependence of interactions on the angle between spin vector and space direction.

To investigate the defect dynamics, we initialize the system with a $n=\pm 1$ pair of defects 
with $+1$ at the origin and $-1$ at $(x_-,y_-)$, corresponding to an initial orientation field $\theta^i =\tan^{-1}{y^i}/{x^i} - \tan^{-1}(y^i-y_-)/(x^i-x_-) + \phi$ at the $i$th lattice point. We choose the overall phase $\phi=0$, so that the background far-field spin configuration points to the \emph{right}. 
Our initial defect configurations are shown in Fig.~\ref{fig:defect-field}. 

\begin{figure*}
\centering
\includegraphics[scale=0.5]{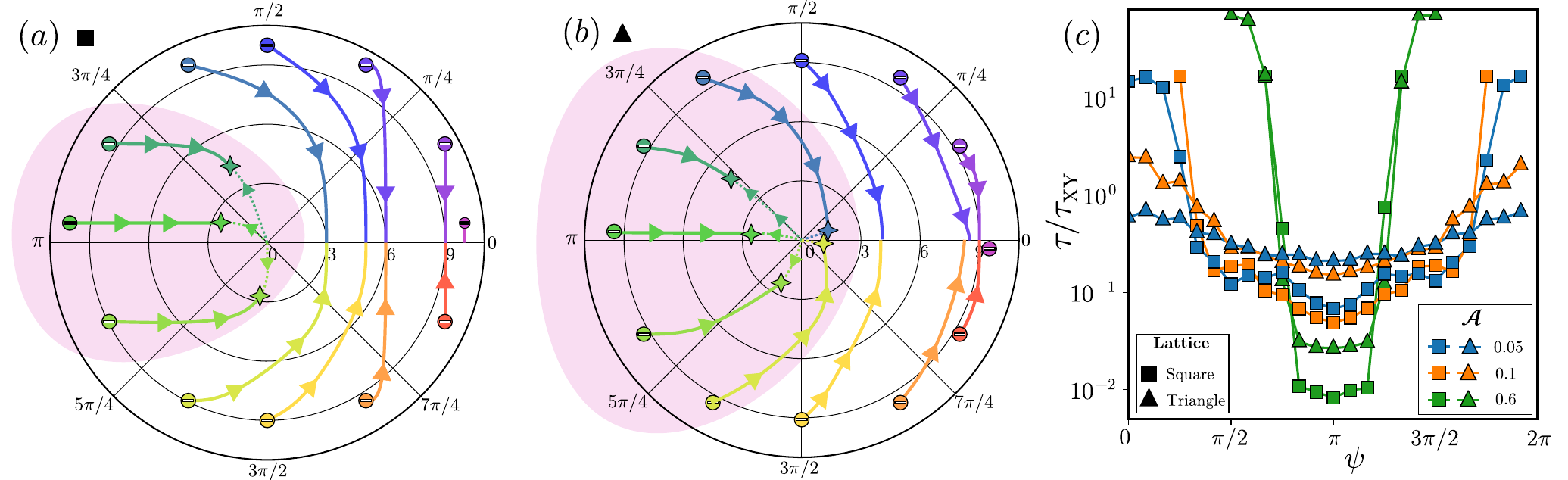}
\caption{
{Quantification of anisotropic defect dynamics. The pink shaded region in {(a)} and {(b)} depicts the angular range of $\psi$ for which topological defects recombine within our simulation time for $\mathcal{A}=0.6$ and $T=0.01$ (such that $r_c$ is of the order of the lattice spacing) in square and triangular lattice respectively. We show representative trajectories of defect pairs with the $+1$ defect placed at the origin and the $-1$ defect placed at $r_0=10$ at various angles, with $L=64$. The smoothed trajectories of $-1$ defects are depicted with solid lines while those of the $+1$ defects with dashed lines. The positions of their annihilation (when they do annihilate) are marked by stars. The subfigure {(c)} shows the pair lifetime $\tau$ scaled by the lifetime of a $\pm 1$ pair with the same initial $r_0$ in a system with $\mathcal{A}=0$, as a function of $\psi$, for three different values of $\mathcal{A}$. The plots cut off when $\tau$ is larger than the simulation time. This demonstrates that the angular range in which defects merge increases with decreasing $\mathcal{A}$.}
}
\label{fig:defect-trajectories}
\end{figure*}
\begin{figure}[h!]
\centering
\includegraphics[scale=0.3]{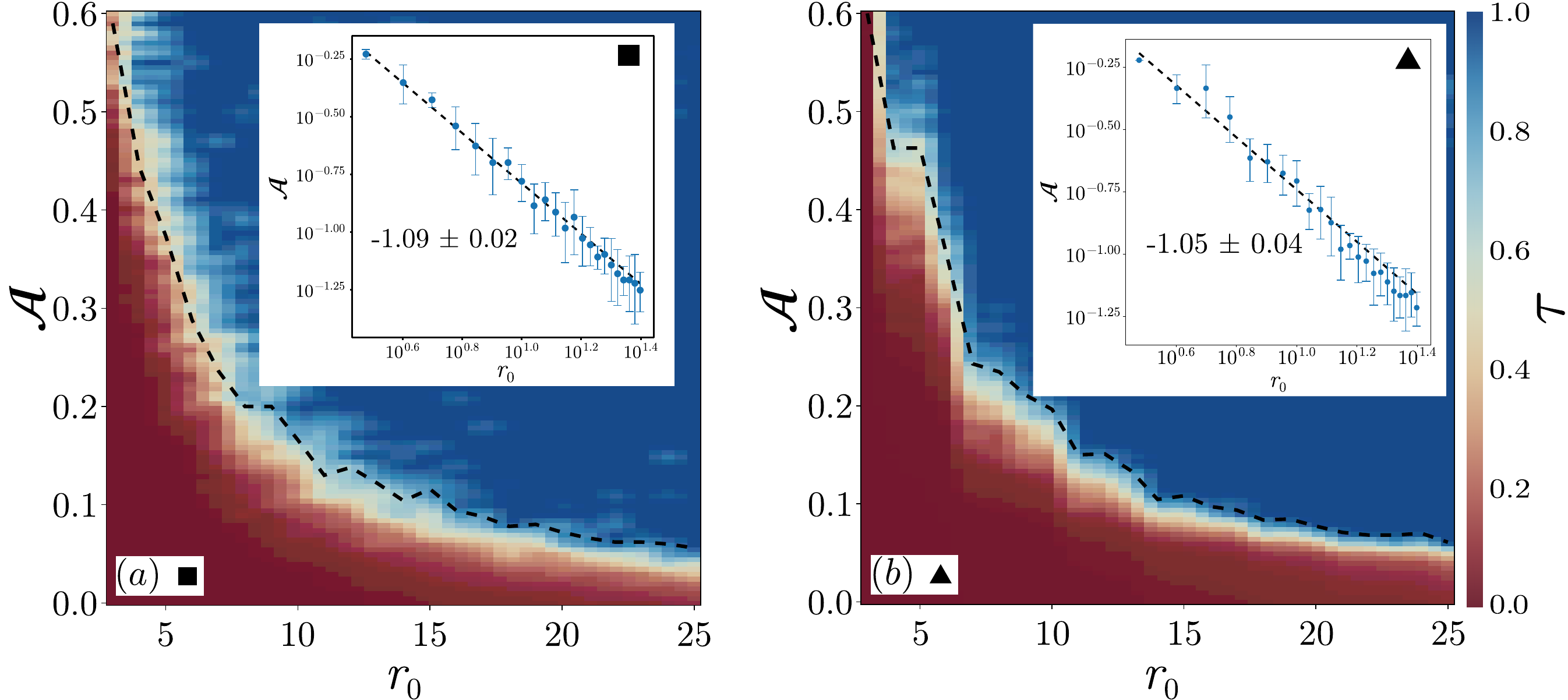}
\caption{{The heatmap shows the recombination time $\tau$ for a $\pm 1$ defect pair with initial $\psi=0$ and separation $r_0$ for square {(a)} and triangular {(b)} lattices. $\tau$ is normalised to $1$ by the total simulation time ($10^7$ timesteps) and, therefore, a recombination time of $1$ means that the defects did not recombine within the simulation time. The black, dashed line is the boundary in the $r_0-\mathcal{A}$ plane between the configurations that become free of defects and those that do not.}
Inset: {{The blue points are {extracted} from the boundary between the red and blue regions and averaged over $5$ different noise realisations for the square lattice and $11$ different noise realisations for the triangular lattice}. The error bars correspond to the variations due to different noise realisations. Black dashed lines are least-square fits to these points which show the existence of a cloaking lengthscale $r_c \propto 1/\mathcal{A}$.}
}
\label{fig:lifetime-heatmap}
\end{figure}

For $\mathcal{A}=0$, the invariance of the model under independent space and spin rotations implies that defects interact logarithmically irrespective of their shape or position in the physical space. When $\mathcal{A}\neq 0$, in contrast, the defect interactions depend on $\psi${$=\tan^{-1}(y\_/x\_)$} as shown by the snapshots in Fig.~\ref{fig:defect-configs}{a-c} for $\pm 1$ pairs with different starting $\psi$. Importantly, Fig.~\ref{fig:defect-trajectories}, which shows the trajectories of $-1$ defects initially placed at different angles with respect to the $+1$ defect, demonstrates that the time required for a $\pm 1$ pair to annihilate depends strongly on $\psi$: the pink shading marks the region in which the time $\tau$ for a $-1$ defect to merge with the $+1$ defect is faster than for the equilibrium XY case, i.e., for $\mathcal{A}=0$ and decreases with $\mathcal{A}$. Outside this region, $-1$ rotates around $+1$ until it reaches $\psi=0$, after which merger takes place with a $\tau$ that increases with $\mathcal{A}$ (see Fig.~\ref{fig:defect-configs}b). {Both the lengthscale at which the speed of $-1$ defects decrease significantly in front of the $+1$ defect and the anisotropy of defect recombination times may be heuristically understood on the basis of the continuum dynamical equation for the polarisation \eqref{cont_sq}. This equation has a natural lengthscale $r_c\equiv J/\mathcal{A}$. Extracting a ``cloaking length'' from the boundary of the parameter regime in which the pair lifetime is larger than the simulation time
in Fig.~\ref{fig:lifetime-heatmap}, we find that, like $r_c$, it is inversely proportional to $\mathcal{A}$ (see inset Fig.~\ref{fig:lifetime-heatmap}). Further, $r_c$ is a ``signed'' lengthscale: depending on the relative orientation of the wavevector and the ordering direction, the relative signs of $\partial_x\theta$ and $\nabla^2\theta$ (terms from which this lengthscale is constructed) change. Therefore, depending on the orientation of the $\pm 1$ pair with respect to the far-field ordering direction, we expect either an enhanced or a reduced merger time.}

\begin{figure}[h!]
\centering
\includegraphics[scale=0.15]{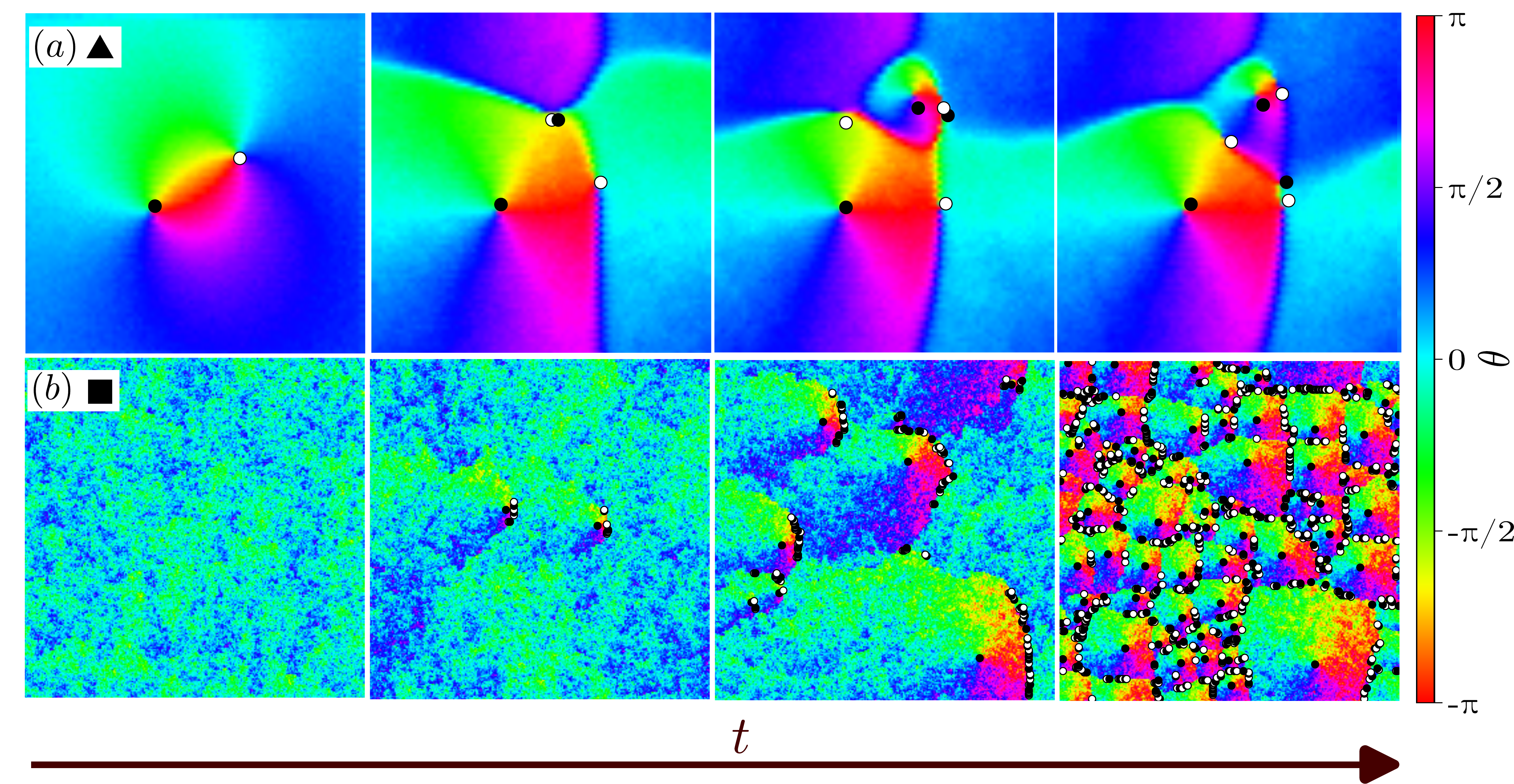}
\caption{{(a) Defect nucleation in regions of strong deformation on a triangular lattice of size $L=80$, and $\mathcal{A}=0.5$ and $T=0.005$. Two additional $\pm 1$ pairs are nucleated in region in which the orientation field has large gradients because of the presence of a $\pm 1$ defect pair. The original $-1$ defect finally merges with one of the newly created $+1$ defects. Also see Supplementary Video SV4. (b) Destruction of an initially defect-free ordered state by the proliferation of defect pairs on a square lattice with $L=256$, $\mathcal{A}=2.0$, and $T=0.4$. Also see Supplementary Video SV5.}
}
\label{fig:defect_nucleation}
\end{figure}
We believe that our observation could provide an explanation for the aster explosion observed in \cite{Besse2022} although lattice effects, e.g., a difference in the shaded recombination regions for the square and triangular cases  (Fig.~\ref{fig:defect-trajectories}a and b), do enter our study.
Lattice anisotropy also affects 
the far-field shape of topological defects\footnote{The shape of the \textit{core} of the defect is nonuniversal and strongly model-dependent even in equilibrium systems (both isotropic and anisotropic). This further implies that defect mobility is not universal and cannot be obtained from the dynamics of spin waves; only the far field can be accounted for by such a theory.} in the hydrodynamic limit. Therefore, the defect interactions and dynamics obtained in our model (and those of \cite{Rouzaire2024, Loos2022a, Loos2024}) are not expected to be equivalent to those of a Malthusian flock in isotropic space. However, lattice anisotropy usually stabilises order and makes unbinding of defects \emph{less} likely. Therefore, if we find an instability mechanism involving defects in this lattice model, it is likely to still operate in the isotropic model. Indeed we observe that as the $-1$ slows down it creates large-scale, wall-like loci of strong deformation encircling the aster, on which new defect pairs are nucleated ({see Supplementary Video (SV4) and Fig. \ref{fig:defect_nucleation}. a}) as reported in \cite{Besse2022}. {In common with \cite{Besse2022} and \cite{Solon_comm} (the latter using the model in Eq. \eqref{eq:lattice-model}), we also find that a large number of defect pairs are nucleated in an initially ordered state, with no topological defects, at long times, and polar order is destroyed via this aster explosion (see Supplementary Video (SV5) and Fig. \ref{fig:defect_nucleation}. b).}

\begin{figure}[h]
\centering
\includegraphics[scale=0.4]{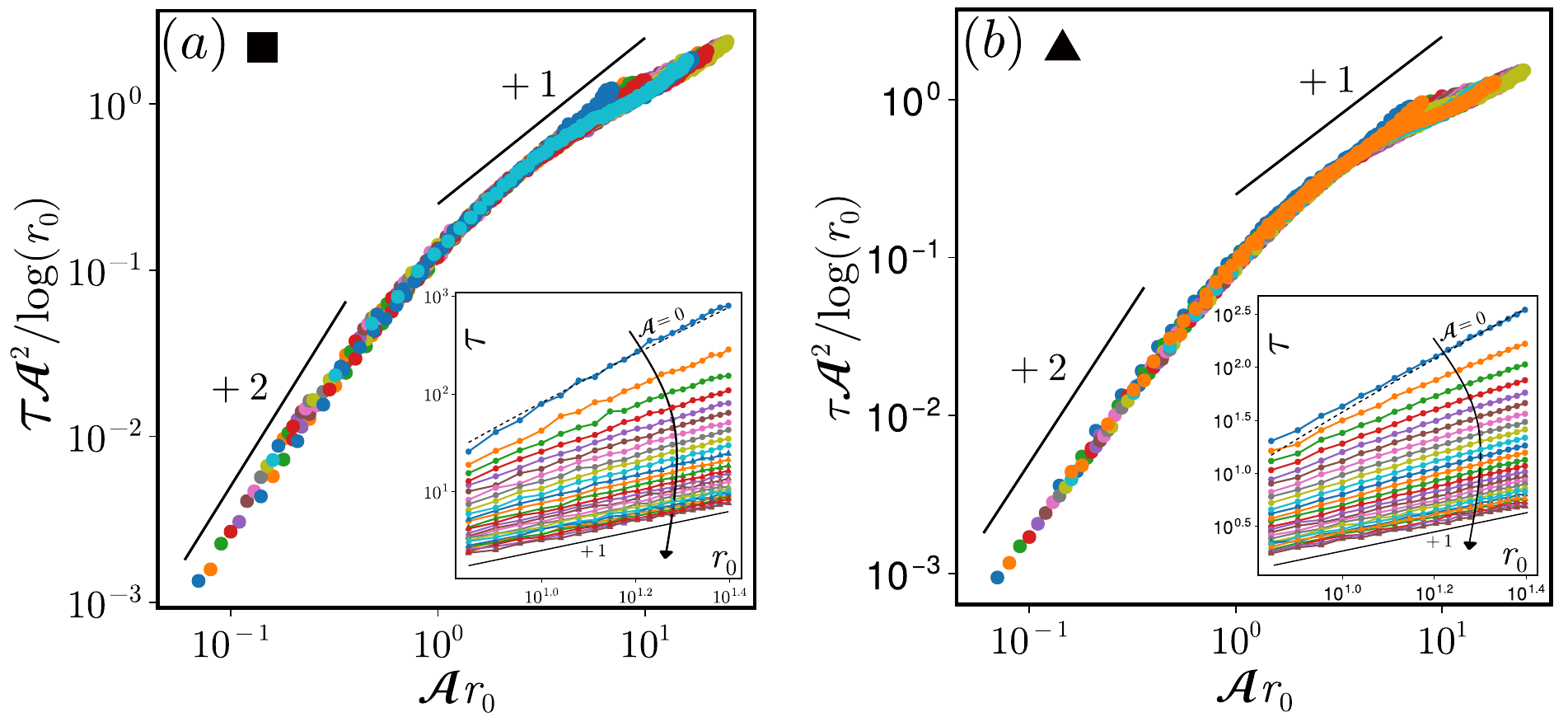}
\caption{
Scaled pair lifetime (data from the inset) for the defects placed at $\psi=\pi$ for {(a)} square and {(b)} triangular lattice. Here box size $L=80$ at temperature $T=0.05$. The initial defect separation were taken in the range $r_0 \in \{5,25\}$ and non-reciprocity $\mathcal{A} \in \{0,1\}$ respectively. Given the $\log$ corrections, pair lifetime $\tau$ shows a crossover from XY dominated dynamics ($+2$) to nonreciprocal dominated ($+1$). Inset: variation of pair lifetime as a function of $r_0$ and different values of $\mathcal{A}$.
The dashed line corresponds to the analytical solution \cite{Denniston,Yurke1993,Radzihovsky2015} $\tau \propto r^2\log(r)$ for XY ($\mathcal{A} =0 $) case. For larger values of non-reciprocity the pair lifetime $\tau \propto r/\mathcal{A}$.
}
\label{fig:collapse}
\end{figure}

We now discuss the timescale $\tau$ of defect merger (when the defects do merge). 
When $\mathcal{A}=0$, equilibrium $XY$-model dynamics prevails and the inverse timescale associated with recombination is $\tau^{-1}\sim 1/r_0^2 \log(r_0)$ \cite{Denniston,Yurke1993,Radzihovsky2015}, where $r_0$ is the initial defect separation. If we assume that the effect of $\mathcal{A}$ is to ``flow'' the $-1$ defect into the $+1$ defect, the recombination rate with $\mathcal{A}\neq 0$ should have an additional contribution $\tau^{-1}\sim \mathcal{A}/r_0$. This implies that $\tau\mathcal{A}^2$ should vary as $\mathcal{A}r_0^2\log r_0$ at small $r_0$ and as $\mathcal{A}r_0$ at large $r_0$. In Fig.~\ref{fig:collapse} we demonstrate that this simple interpolation is borne out by our simulations.

\section{Discussion}
{In this article, we discuss lattice models of active planar rotors whose interactions carry a nonreciprocality governed by their orientation with respect to their relative positions \cite{dadhichi}. We demonstrate that topological defect interactions are highly anisotropic with respect to the mean ordering direction, with $\pm 1$ defect pairs in certain configurations effectively shielded from each other, creating long-lasting wall-like deformations along which new defect pairs are preferentially nucleated. We offer numerical evidence that this process leads to the destruction of the ordered phase via an aster apocalypse similar to \cite{Besse2022}. We show, however, that the long-wavelength theory of our model (and those of \cite{Rouzaire2024,Loos2022a,Loos2024}) always retains traces of lattice anisotropy  
and is therefore not that of the Malthusian flock \cite{Toner2012}. 
Nonetheless, since lattice anisotropy is likely to stabilise rather than destabilise the ordered phase, we argue that this phenomenon, which is consistent with the discussion in Ref. \cite{Besse2022} should be more pronounced in flocks in the continuum.}

{Important directions for the future include eliminating the effects of the lattice anisotropy in our model as well as in models such as \cite{Rouzaire2024}, by using strategies outlined in \cite{CLT1,CLT2} for instance, and using this to calculate the \emph{universal} scaling of static and dynamic correlation fluctuations of angular fluctuations. As has been recently discussed \cite{AM_malthus}, these quantities \emph{cannot} be analytically calculated in two dimensions and their earlier predicted values are incorrect. It would also be worthwhile to use the isotropised lattice model to examine the dynamics of topological defects and, in particular, whether they generically proliferate for any nonzero non-reciprocity leading to the destruction of the ordered state.} 

\begin{acknowledgements}
{PP acknowledges discussions with Sumantra Sarkar, Navdeep Rana, and Vikash Pandey, and SR and AM 
with Demian Levis and Sarah Loos.}
AM acknowledges the support of ANR through the grant PSAM. AM and SR thank the Isaac Newton Institute for Mathematical Sciences for support and hospitality during the program ``Anti-diffusive dynamics: from sub-cellular to astrophysical scales'' (EPSRC grant no. EP/R014604/1), and SR additionally for a Rothschild Distinguished Visiting Fellowship. This research was supported in part by grant NSF PHY-2309135 to the Kavli Institute for Theoretical Physics. AM acknowledges a TALENT fellowship awarded by CY Cergy Paris Universit\'e, and SR a JC Bose Fellowship of the ANRF, India. 
\end{acknowledgements}
\appendix
\section{\label{A:angle_field_triangular}
Expressions for the angle field equation for triangular lattices}
In this appendix, we display the continuum equations for $\theta$ for a triangular lattice. For the continuum equations for $\theta$ on a triangular lattice for the model of \cite{Rouzaire2024}, we use
\eqref{expan} to expand $\sum _{j\in~\mathcal{N}(i)}Je^{\sigma\hbf{p}^i\cdot\hbf{R}^{ij}}\hbf{p}^i \times \hbf{p}^j\times\hbf{p}^i$ to second order in gradients and fifth order in $\sigma$:
\begin{multline}
   \left(\sum _{j\in~\mathcal{N}(i)} Je^{\sigma\hbf{p}^i\cdot\hbf{R}^{ij}}\hbf{p}^i \times \hbf{p}^j\times\hbf{p}^i\right)_\alpha\approx J\delta_{\alpha\beta}^T\sum_a\Bigg[\frac{1}{2}{\bf e}_\mu^a {\bf e}_\gamma^a\partial_\mu\partial_{\gamma}\hbf{p}_\beta({\bf x}_i,t)+\sigma\hbf{p}_\mu({\bf x}_i,t){\bf e}_\mu^a {\bf e}_\gamma^a\partial_{\gamma}\hbf{p}_\beta({\bf x}_i,t)\\+\frac{\sigma^2}{4}\hbf{p}_\mu({\bf x}_i,t)\hbf{p}_\nu({\bf x}_i,t){\bf e}_\mu^a{\bf e}_\nu^a{\bf e}_\lambda^a {\bf e}_\gamma^a\partial_\lambda\partial_{\gamma}\hbf{p}_\beta({\bf x}_i,t)+\frac{\sigma^3}{6}\hbf{p}_\mu({\bf x}_i,t)\hbf{p}_\nu({\bf x}_i,t)\hbf{p}_\lambda({\bf x}_i,t){\bf e}_\mu^a{\bf e}_\nu^a{\bf e}_\lambda^a {\bf e}_\gamma^a\partial_{\gamma}\hbf{p}_\beta({\bf x}_i,t)\\
   +\frac{\sigma^4}{48}\hbf{p}_\mu({\bf x}_i,t)\hbf{p}_\nu({\bf x}_i,t)\hbf{p}_\chi({\bf x}_i,t)\hbf{p}_\xi({\bf x}_i,t){\bf e}_\xi^a{\bf e}_\nu^a{\bf e}_\chi^a{\bf e}_\mu^a{\bf e}_\nu^a{\bf e}_\lambda^a {\bf e}_\gamma^a\partial_\lambda\partial_{\gamma}\hbf{p}_\beta({\bf x}_i,t)\\+\frac{\sigma^5}{120}\hbf{p}_\mu({\bf x}_i,t)\hbf{p}_\nu({\bf x}_i,t)\hbf{p}_\lambda({\bf x}_i,t)\hbf{p}_\chi({\bf x}_i,t)\hbf{p}_\xi({\bf x}_i,t){\bf e}_\xi^a{\bf e}_\chi^a{\bf e}_\mu^a{\bf e}_\nu^a{\bf e}_\lambda^a {\bf e}_\gamma^a\partial_{\gamma}\hbf{p}_\beta({\bf x}_i,t)\Bigg]\,.
\end{multline}

{
Using this and the fact that $\hbf{p}({\bf x},t)\equiv\{\cos\theta({\bf x},t),\sin\theta({\bf x},t)\}$, we obtain the equation for $\theta$ on a  triangular lattice:
\begin{multline}
\partial_t\theta=3J\Bigg\{\frac{1}{2}\nabla^2\theta+\sigma(\cos\theta\partial_x\theta+\sin\theta\partial_y\theta)+\frac{\sigma^2}{16}[(\cos2\theta+2)\partial_x^2\theta+2\sin2\theta\partial_x\partial_y\theta+(2-\cos2\theta)\partial_y^2\theta]\\+\frac{\sigma^3}{8}(\cos\theta\partial_x\theta+\sin\theta\partial_y\theta)+\frac{\sigma^4}{768}[(6+4\cos2\theta+\cos4\theta)\partial_x^2\theta+2(4\sin2\theta-\sin4\theta)\partial_x\partial_y\theta\\+(6-4\cos2\theta-\cos4\theta)\partial_y^2\theta]+\frac{\sigma^5}{1920}[(10\sin\theta-\sin5\theta)\partial_x\theta+(10\cos\theta+\cos5\theta)\partial_y\theta]\Bigg\}+\sqrt{2T}\xi_\theta\,.
\label{Levis_cont_tri}
\end{multline}
This equation is invariant under \eqref{rotinv} up to $\mathcal{O}(\sigma^3)$ (despite the rather complicated-looking form of the term $\propto\sigma^2$, it is anisotropic diffusion). However, the term $\propto\sigma^4$, when expanded to $\mathcal{O}(\theta^3)$ has a nonlinear term $\theta^2(\partial_y^2-\partial_x^2)\theta$, which we have argued places this model in the active clock model universality class \cite{Toner_clock}.}

We now expand \emph{our} model to $\mathcal{O}(\nabla^5)$ on a triangular lattice; the angle field dynamics (the triangular lattice version of Eq. \eqref{cont_sq4}) then reads
\begin{multline}
\partial_t\theta=3\Bigg[\mathcal{A}(\cos\theta\partial_x\theta+\sin\theta\partial_y\theta)+\frac{J}{2}\nabla^2\theta+\frac{3\mathcal{A}}{24}\{\cos\theta[\partial_x^3\theta+\partial_x\partial_y^2\theta-(\partial_x\theta)^3-\partial_x\theta(\partial_y\theta)^2]+\sin\theta[\partial_y^3\theta+\partial_x^2\partial_y\theta-(\partial_y\theta)^3-(\partial_x\theta)^2\partial_y\theta]\}\\+\frac{J}{32}[\partial_x^4\theta+\partial_y^4\theta+2\partial_x^2\partial_y^2\theta-8\partial_x\theta\partial_y\theta\partial_x\partial_y\theta-2(\partial_y\theta)^2(3\partial_y^2\theta+\partial_x^2\theta)-2(\partial_x\theta)^2(3\partial_x^2\theta+\partial_y^2\theta)]\\+
\frac{\mathcal{A}}{1920}\Bigg(\cos\theta\{15(\partial_y\theta)^4\partial_x\theta+15\partial_x\partial_y^4\theta-60\partial_y\theta[\partial_x\partial_y\theta(3\partial_y^2\theta+\partial_x^2\theta)+\partial_x\theta\partial_y\nabla^2\theta]+10(\partial_y\theta)^2[(\partial_x\theta)^3-9\partial_x\partial_y^2\theta-\partial_x^3\theta]\\+\partial_x\theta[11(\partial_x\theta)^4-60(\partial_x\partial_y\theta)^2-45(\partial_y^2\theta)^2-30\partial_x^2\theta\partial_y^2\theta-165(\partial_x^2\theta)^2-10\partial_x\theta(3\partial_y^2\partial_x\theta+11\partial_x^3\theta)]+10\partial_x^3\partial_y^2\theta+11\partial_x^5\theta\}\\+\sin\theta\{9(\partial_y\theta)^5+9\partial_y^5\theta+30(\partial_y\theta)^3(\partial_x\theta)^2-90(\partial_y\theta)^2\partial_y\nabla^2\theta-5\partial_y\theta[-(\partial_x\theta)^4+36(\partial_x\partial_y\theta)^2+3(3\partial_y^2\theta+\partial_x^2\theta)^2+4\partial_x\theta(9\partial_x\partial_y^2\theta+\partial_x^3\theta)]\\-30\partial_x\theta[2\partial_x\partial_y\theta(3\partial_y^2\theta+\partial_x^2\theta)+\partial_x\theta\partial_y\nabla^2\theta]+30\partial_x^2\partial_y^3\theta+5\partial_y\partial_x^4\theta\}\Bigg)\Bigg]
+\sqrt{2T}\xi_\theta\,.
\label{cont_tri4}
\end{multline}
This equation is invariant under the transformation in \eqref{rotinv} only up to $\mathcal{O}(\nabla^4)$ (though all terms that do not depend on $\mathcal{A}$ must be symmetric under $\theta\to \theta+\theta_0$ to arbitrary order in gradients); we checked that the term $\propto\mathcal{A}$ in the large round brackets is not invariant under Eq. \eqref{rotinv}.

\section{{Additional measurements of the properties of the ordered phase in our model and in the model of \cite{Rouzaire2024}}}
\label{app:struct}
\begin{figure}[h]
\centering
\includegraphics[scale=0.35]{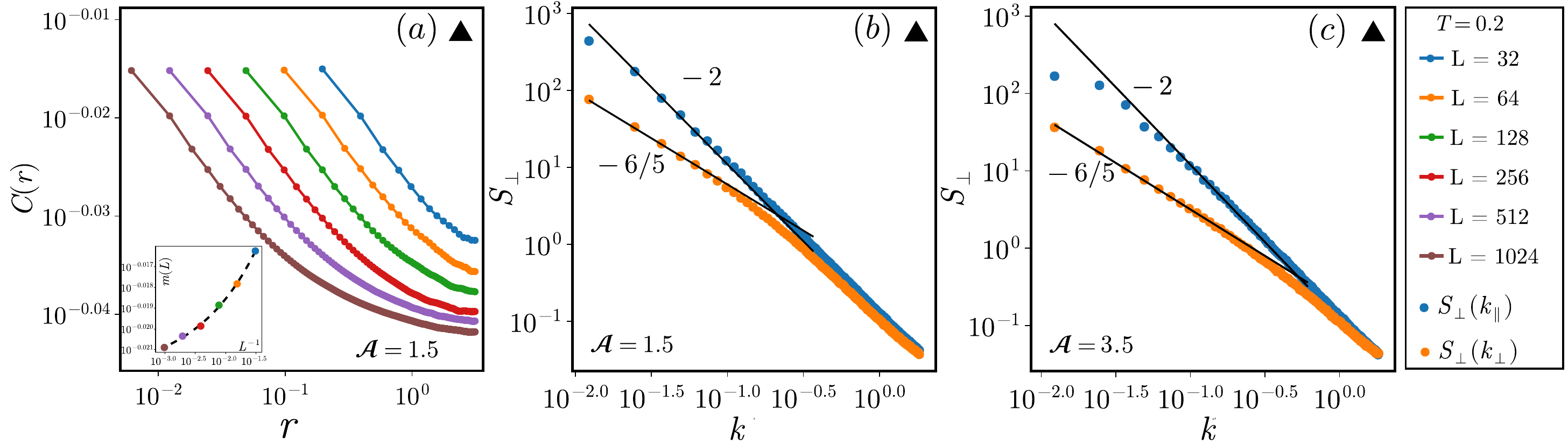}
\caption{
{Statistical properties of the ordered phase on a triangular lattice. (a) Two-point correlation function $C(r)$ for various system sizes $L$, with distance $r$ in units of $L/2\pi$, approaches a constant nonzero value for large $r$, indicating long-range order. {Averaging is carried out over five independent noise realizations.} Inset shows magnetization $m(L)\equiv \sqrt{ C(r=L)}$ as a function of $L^{-1}$. (b) Measured structure factors $S_{\perp}$ as functions of $k_\parallel$ and $k_\perp$ for $L=512$. The analysis is performed at reduced temperature $T=0.2$ and nonreciprocity strength $\mathcal{A}=1.5$, with time averaging over four independent noise realizations.}
{(c) Measured structure factors $S_{\perp}$ as functions of $k_\parallel$ and $k_\perp$ for $L=512$, with time averaging over four independent noise realizations for $\mathcal{A}=3.5$. The flattening out of the structure factor at small wavenumbers was not very clear for $\mathcal{A}=2.5$ in the triangular lattice, in the wavenumber regime we examine; therefore, we display the result for $\mathcal{A}=3.5$ where we observe that the scaling of the structure factor becomes independent of $k_\parallel$ at small $k_\parallel$.}
}
\label{fig:STF_ANRI_2.5}
\end{figure}



\begin{figure}[h]
\centering
\includegraphics[scale=0.4]{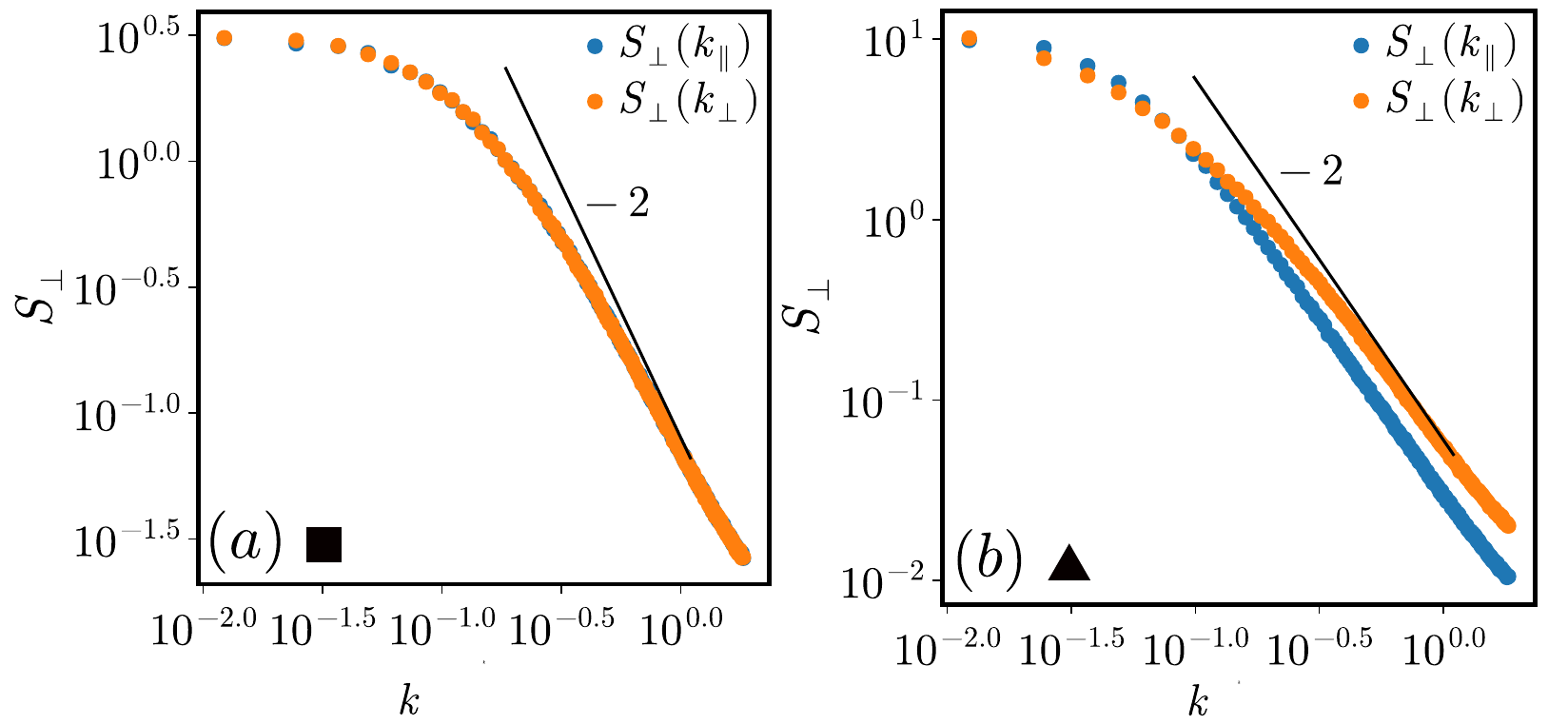}
\caption{{Measured structure factors $S_{\perp}$ as functions of $k_\parallel$ and $k_\perp$ for $L=512$ for {(a)} square and {(b)} triangular lattice for the model in \cite{Rouzaire2024}. The analysis is performed at reduced temperature $T=0.2$ and $\sigma=2.5$ with time-averaging over five noise realisations. The flattening of $S_\perp$ at small wavenumbers clearly implies a {nonzero} relaxation rate of angular fluctuations in this limit due to the presence of a term $\propto\theta$ in this case.}
}
\label{fig:STF_ROUZ_A_2.5}
\end{figure}
\newpage
\section{Simulation details} 
We numerically integrate the equation (\ref{eq:lattice-model})
with nearest neighbour couplings $J^{ij}\equiv J+\mathcal{A}\hbf{p}^i\cdot\hbf{R}^{ij}$
in two dimensions on square and triangular lattices of linear size $L$, with $J=1$, using the forward Euler scheme and periodic boundary conditions. 
Our control parameters are the temperature $T$ and the non-reciprocity $\mathcal{A}$. 
$\zeta^i$ is unit-strength zero-mean Gaussian white noise. We choose a time step of $\Delta t = 0.001$, and run long enough to reach stationarity and acquire sufficient steady-state data.
\section{Details of supplementary videos}
\begin{itemize}
    \item[{\bf SV1}]: \texttt{Defect\_screening.mkv}: {The dynamics} of a {$\pm 1$} defect pair {with an initial separation $r_0=10$ and $\psi=0$, on an $L=64$ square lattice.}
    The control parameters are $\mathcal{A}=0.6$, and $T=0.01$.
    \item[{\bf SV2}]: \texttt{Defect\_rotation.mkv}: 
    {The dynamics of a $\pm 1$ defect pair with an initial separation $r_0=10$ and $\psi=7\pi/12$, on an $L=64$ square lattice. The control parameters are $\mathcal{A}=0.6$, and $T=0.01$.}
    \item[{\bf SV3}]: \texttt{Enhanced\_annihilation.mkv}: {The dynamics of a $\pm 1$ defect pair with an initial separation $r_0=10$ and $\psi=\pi$, on an $L=64$ square lattice. The control parameters are $\mathcal{A}=0.6$, and $T=0.01$.}
    \item[{\bf SV4}]: \texttt{Defect\_nucleation.mkv}: 
    {The nucleation of new defect pairs during the time evolution of a $\pm 1$ defect pair with an initial separation $r_0=25$ and $\psi=\pi/6$, on an $L=80$ triangular lattice. The control parameters are $\mathcal{A}=0.5$, and $T=0.005$.}
    \item[{\bf SV5}]: \texttt{Aster\_apocalypse.mkv}: {The evolution of an initially ordered state on a $L=256$ square lattice. The control parameters are $\mathcal{A}=2.0$ and $T=0.4$.}
\end{itemize}

\bibliography{refs/NRIXY-refs}
\end{document}